\documentstyle[a4,12pt,epsf]{article}
\epsfverbosetrue
\newcommand{\figurebox}[2]{\fbox{\vbox to#2in{\hbox to #1in{\hfil} \vfil}}}
\newcommand{\err}[2]{{\footnotesize {$\;\begin{array}{@{}l@{}}
			  +\makebox[1.3em][r]{#1} \\[-0.4em]
			  -\makebox[1.3em][r]{#2}
			\end{array}$}}}
\newcommand{\pseudo}{{PS}}
\newcommand{\beq}{\begin{equation}}
\newcommand{\eeq}{\end{equation}}

\newcommand{\zero}[1]{\multicolumn{1}{#1}{0.0}}
\begin{document}

\begin{titlepage}

\begin{flushright}
Edinburgh Preprint: 92/513 \\
Liverpool Preprint: LTH 302 \\
Southampton Preprint: SHEP 92/93-9 \\
\end{flushright}
\vspace{5mm}
\begin{center}
{\Huge Gauge-Invariant Smearing and Matrix Correlators using Wilson
Fermions at $\beta=6.2$}\\[10mm] {\large\it UKQCD
Collaboration}\\[3mm]

{\bf C.R.~Allton\footnote{Present Address: Dipartimento di Fisica,
Universit\`{a} di Roma {\it La Sapienza}, 00185 Roma, Italy},
C.T.~Sachrajda}\\ Physics Department, The University, Southampton
SO9~5NH, UK

{\bf R.M.~Baxter, S.P.~Booth, K.C.~Bowler, S.~Collins, D.S.~Henty,
R.D.~Kenway,
B.J.~Pendleton, D.G.~Richards, J.N.~Simone,
A.D.~Simpson, B.E.~Wilkes}\\
Department of Physics, The University of Edinburgh, Edinburgh EH9~3JZ,
Scotland

{\bf C.~Michael}\\
DAMTP, University of Liverpool, Liverpool L69~3BX, UK

\end{center}
\vspace{5mm}

\begin{abstract}

We present an investigation of gauge-invariant smearing for Wilson
fermions in quenched lattice QCD on a $24^3 \times 48$ lattice at
$\beta = 6.2$.  We demonstrate a smearing algorithm that allows a
substantial improvement in the determination of the baryon spectrum
obtained using propagators smeared at both source and sink, at only a
small computational cost.  We investigate the matrix of correlators
constructed from local and smeared operators, and are able to expose
excited states of both the mesons and baryons.

\end{abstract}

\end{titlepage}

\section{Introduction}\label{introduction}
	The use of ``smeared'' interpolating fields to expose more
clearly the leading behaviour of Euclidean correlators has become
standard practice in lattice gauge simulations.  The extended sources
first suggested in~\cite{kenway84,billoire85} comprise a sum of delta
functions on the source timeslice and therefore are not
gauge-covariant.  These sources give an improved signal for small
Euclidean times, but lead to increased statistical noise at large
times.  The noise problem can be cured by fixing to Coulomb gauge on
the source timeslice, and the resulting ``cube''~\cite{APE88} and
``wall''~\cite{kilcup89} sources, and variants
thereof~\cite{degrand91,butler91}, have been used in many areas of
lattice QCD phenomenology.  A more ambitious approach is to measure
Coulomb-gauge wave functions, and to use these as sources to improve
the overlap with the ground state~\cite{degrand92,elkhadra92,fnal92}.

When fixing to a smooth gauge such as Coulomb or Landau gauge, one
must be careful that ambiguities in the gauge-fixing
condition~\cite{parrinello91} do not affect the final result.  In this
paper we avoid the gauge fixing problem through the use of
gauge-covariant sources~\cite{guesken90,eichten90}.

The plan of this paper is as follows.  We begin by comparing effective
masses obtained from local sinks with those from Wuppertal
scalar-propagator smeared sinks~\cite{guesken90}.  We introduce a
variant of the alternative iterative Wuppertal scheme which enables us
to achieve similar results to scalar propagator smearing, but at a
much lower cost in computer time.  Using this scheme, we perform
detailed measurements of the hadron spectrum and the pseudoscalar
decay constant using local propagators (LL), propagators smeared at
the source only (LS), at the sink only (SL), and at both source and
sink (SS).  Finally, we investigate the $2
\times 2$ matrix of correlators formed from propagators computed
with the four source-sink combinations, and by diagonalising this
matrix are able to isolate the first excited state in the various
meson and baryon channels.

\section{Euclidean Correlators}
	In lattice field theory, masses and matrix elements are
obtained from the asymptotic exponential decay of two-point
correlation functions of the form
\begin{equation}
c_{{\cal O}\tilde{{\cal O}}}(t)
  = \sum_{\vec{x}} \langle {\cal O}(\vec{x},t)
    \tilde{{\cal O}}^{\dagger} (\vec{0},0) \rangle e^{-i \vec{p}\cdot \vec{x}}
    \label{c-of-t}
\end{equation}
	where the operator $\tilde{{\cal O}}^\dagger$ at the source
$(\vec{0},0)$ has the appropriate quantum numbers to create the hadron
$h$ of interest.  This hadron is destroyed at the sink by an
appropriate operator ${\cal O}(\vec{x},t)$.  Inserting a complete set
of eigenstates of the Hamiltonian, we find
\begin{equation}
c_{{\cal O}\tilde{{\cal O}}}(t) = \frac{\langle 0 | {\cal O}(0) |
h_0(\vec{p}) \rangle
\langle h_0(\vec{p}) | \tilde{{\cal O}}^{\dagger} (0) |0 \rangle}
{2 E_0(\vec{p})} \,e^{-E_0(\vec{p}) t} \, A(t)
	\label{spectral-rep}
\end{equation}
where the correction $A(t)$ to the asymptotic behaviour is
\begin{equation}
A(t) = 1 + \sum_{n=1}^\infty	\frac{E_0(\vec{p})}{E_n(\vec{p})}
    \frac{ \langle 0 | {\cal O}(0) | h_n(\vec{p}) \rangle \,
	    \langle h_n(\vec{p}) | \tilde{{\cal O}}^{\dagger} (0) |0 \rangle
	 }
	 { \langle 0 | {\cal O}(0) | h_0(\vec{p}) \rangle \,
            \langle h_0(\vec{p}) | \tilde{{\cal O}}^{\dagger} (0) |0 \rangle
         } \,
		e^{-(E_n(\vec{p}) - E_0(\vec{p}))t} .
\label{higher_states}
\end{equation}
$|h_0(\vec{p})\rangle$ is the lightest hadron state with 3-momentum
$\vec{p}$ that can be created by $\tilde{\cal O}^\dagger$, and
$|h_n(\vec{p})\rangle$ is its $n$th excited state.  With a suitable
choice of ${\cal O}$ and $\tilde{{\cal O}}$, we hope to enhance the
matrix elements between the vacuum and the lightest state, so that the
coefficients of the non-leading exponentials of
Equation~(\ref{higher_states}) are reduced, and the asymptotic
behaviour ($A(t) \to 1$) is reached for small $t$, where statistical
errors are small.

\section{Gauge-Invariant Smearing}
\subsection{Computational Details}
	Our simulations were performed on a $24^3 \times48 $ lattice
at $\beta=6.2$ using quenched Wilson fermions, with a hopping
parameter $\kappa=0.152$. This corresponds to a pseudoscalar meson
mass of approximately $600$ MeV if we set the scale from the string
tension ($a^{-1}\sim2.73(5)$ GeV~\cite{hadrons_letter}), or from
$m_\rho$.  The configurations were generated on an Meiko i860
Computing Surface, and the propagators on a Thinking Machines CM-200,
both at Edinburgh.

\subsection{Wuppertal Smearing}
The Wuppertal scalar-propagator smeared source~\cite{guesken90},
$S(\vec{x'},\vec{0})$, is given by the solution of the three
dimensional Klein-Gordon equation
\begin{equation}
K(\vec{x},\vec{x'}) S(\vec{x'},\vec{0}) = \delta_{\vec{x},\vec{0}},
\label{kg-green-function}
\end{equation}
where
\begin{equation}
K(\vec{x},\vec{x'}) = \delta_{\vec{x},\vec{x'}} - \kappa_S
\sum_{\mu} \{ \delta_{\vec{x'},\vec{x} - \hat{\mu}}
U^{\dagger}_\mu(\vec{x} - \hat{\mu})
+ \delta_{\vec{x'}, \vec{x} + \hat{\mu}}
U_\mu(\vec{x}) \} .
\end{equation}
This leads to a so-called ``shell-model'' wave
function~\cite{degrand91}, in which each quark is effectively
localised about the origin in a region of radius $r$ controlled by the
scalar hopping parameter $\kappa_{S}$.
The rms radius, $r$, is defined by
\begin{equation}
r^2 = \frac{
 \sum_{\vec{x}} | \vec{x} |^2 | S(\vec{x},\vec{0}) |^2}
{\sum_{\vec{x}} | S(\vec{x},\vec{0}) |^2}.
\end{equation}

For the initial investigation of the dependence of the effective
masses on the smearing radius, quark propagators were computed on an
ensemble of 7 configurations using Wuppertal smeared sources at
$\kappa_S = 0.180$ and $0.184$, corresponding to $r \simeq 2$ and $r
\simeq 4$ respectively. The effective masses of the pseudoscalar, vector,
nucleon and $\Delta$ are shown in Figure
\ref{eff_mass_test}, together with the results for a point source on
the same ensemble of configurations.  The lightest state in each
channel is isolated nearer the origin with increasing smearing radius;
we will investigate below whether this yields an improvement in the
statistical uncertainty on the hadron mass determined from a
correlated fit to the corresponding propagator.
%
%
\begin{figure}[htbp]
\begin{center}
\leavevmode
\epsfysize=250pt
  \epsfbox[20 30 620 600]{masses_LL7.ps}
\epsfysize=250pt
  \epsfbox[20 30 620 600]{masses_SL180.ps}
\end{center}
\begin{center}
\leavevmode
\epsfysize=250pt
  \epsfbox[20 30 620 600]{masses_SL184.ps}
\end{center}
\caption{The effective masses of the pseudoscalar (crosses), vector (diamonds),
nucleon (bursts), and $\Delta$ (squares) on an ensemble of 7
configurations, using propagators calculated with a local source, and
with Wuppertal scalar-propagator smeared sources at $\kappa_S = 0.180$
and $\kappa_S = 0.184$. A local sink is used in each case.}
\label{eff_mass_test}
\end{figure}

The computational overhead of Wuppertal smearing at the source is
relatively small compared with the work required in the computation of
the propagator.  However, the overhead is significant when the
propagator is smeared at the sink, as is generally required to realise
an improvement in the determination of matrix elements, since the
smearing algorithm must be implemented on every timeslice, and for
every spin component.  Indeed, for $\kappa_S = 0.184$, the
computational effort required to smear at the sink is comparable with
that required to compute the propagator, and larger values of the
smearing radius are prohibitively expensive.

\subsection{Jacobi Smearing}

An alternative smeared source can be obtained by solving
Equation~(\ref{kg-green-function}) as a power series in $\kappa_{S}$,
stopping at some finite power $N$.  This can be achieved easily using
Jacobi iteration.  When $\kappa_{S}$ is smaller than some critical
value, the power series converges and one recovers the scalar
propagator.  For sufficiently large $\kappa_{S}$, the series diverges
but provides a valid smeared source for any choice of $N$.  A similar
iterative scheme has been used by the Wuppertal
group~\cite{guesken90,alexandrou91}.

We can construct a smearing function of given radius, $r$, for various
choices of $\{N,\kappa_S\}$.  However, there is a minimum $N$ required
to achieve a particular radius.
Typical source functions with $r
\simeq 4$, for $\{N = 50,\kappa_S = 0.250\}$, $\{N = 90, \kappa_S = 0.190\}$
together with the Wuppertal source function at $\kappa_S = 0.184$, are
shown in Figure
\ref{smearing_shapes}.  For this radius, $N = 50$ is close to
minimal.  The Jacobi source function can be generated in considerably
less computer time than the Wuppertal source function; approximately a
factor of 10 less at this radius.  Furthermore, the truncation of the
Jacobi series suppresses fluctuations in the norm and in $r$ between
configurations; hence there is some additional advantage in taking the
minimum value of $N$.
\begin{figure}[htbp]
\centerline{
  \epsfysize=250pt
  \hspace{30pt}
  \epsffile{jsmear_0.250_50.ps}
  \hspace{-90pt}
  \epsfysize=250pt
  \epsffile{jsmear_0.190_90.ps}
}
\centerline{
  \epsfysize=250pt
  \hspace{30pt}
  \epsffile{wsmear_0.184.ps}
}
\caption{The smearing functions $F(x,y) = \left. \protect\sqrt{ |S|^2} \,
\right|_{z = 0}$, normalised to unit volume,
on a $24^3 \times 48$ lattice using (a) Jacobi smearing with $\kappa_S
= 0.250, N = 50$, (b) Jacobi smearing with $\kappa_S = 0.190, N = 90$
and (c) Wuppertal smearing with $\kappa_S = 0.184$.}
\label{smearing_shapes}
\end{figure}

{}From the spectral representation~(\ref{spectral-rep}), it is easy to
see that, when ${\cal O}=\tilde{{\cal O}}$, correlators obtained from
propagators smeared at the source (LS) and those obtained from
propagators smeared at the sink with local sources (SL) are equal in
the limit that the path integral over the gauge fields is done
exactly.  However, for a finite sample of configurations,
sink-smearing leads to more statistical noise because of fluctuations
in the gauge fields forming the sinks on different timeslices.  We
found that for radii greater than 4, this noise becomes unacceptably
high, though a quantitative comparison of SL and LS results for hadron
masses must await the correlated fits to the data presented below.
Nevertheless, the remainder of this paper will concentrate on results
with $r \simeq 4$ on the same 18 configurations which were studied
in~\cite{hadrons_letter}, using a Jacobi source function with
$\kappa_S = 0.250$, $N = 50$.

To optimise the onset of asymptotic behaviour, we look at propagators
smeared both at the source and the sink (SS).  The effective masses of
the pseudoscalar, vector, nucleon and $\Delta$ obtained using the LL,
LS, SL and SS propagators are shown in Figure~\ref{eff_mass_all}.

%
%
\begin{figure}[htbp]
\begin{center}
\leavevmode
\epsfxsize=250pt
  \epsfbox[20 30 620 600]{masses_LL18.ps}
\epsfxsize=250pt
  \epsfbox[20 30 620 600]{masses_LS.ps}
\end{center}
\begin{center}
\leavevmode
\epsfxsize=250pt
  \epsfbox[20 30 620 600]{masses_SL.ps}
\epsfxsize=250pt
  \epsfbox[20 30 620 600]{masses_SS.ps}
\end{center}
\caption{The effective masses of the pseudoscalar (crosses), vector (diamonds),
nucleon (bursts) and $\Delta$ (squares), computed using the LL, LS, SL
and SS propagators, with a Jacobi smearing algorithm as discussed in
the text.}
\label{eff_mass_all}
\end{figure}

\section{Hadron Masses}
To quantify the preceding discussion, we present in Table
\ref{hadron_masses} the masses in
lattice units of the pseudoscalar, vector, nucleon and $\Delta$ using
the LL, LS, SL and SS propagators, together with the corresponding
time ranges used for the fits. All fits are performed using the full
covariance matrix, with the errors extracted using a bootstrap
analysis \cite{hadrons_paper}. For each of the four source-sink
combinations, the time range was chosen by requiring that the fit be
stable under the removal from the fitting range of the timeslice
closest to the source. The length of the fitting range is constrained
by the number of configurations.  Here we fit to the average of the
appropriate forward and backward propagators, in contrast
to~\cite{hadrons_letter} where we fitted to the forward and backward
propagators independently.

\begin{table}
\begin{center}
\begin{tabular}{|c|cccc|}
\hline
& LL  & SL & LS  & SS  \\
\hline
$m_\pseudo$ & 0.223\err{5}{6} (12 -- 16) & 0.220\err{12}{8} (11 -- 16) &
0.219\err{9}{4} (11 -- 16) & 0.220\err{7}{7} (9 -- 12) \\
$m_V$ & 0.341\err{10}{7} (12 -- 16) & 0.356\err{21}{13} (11 -- 16) &
0.327\err{8}{7} (11 -- 16) & 0.327\err{9}{8} (9 -- 12) \\
$m_N$ & 0.510\err{23}{10} (12 -- 16) & 0.507\err{31}{24} (11 -- 16) &
0.495\err{12}{12} (11 - 16) & 0.495\err{9}{10} (7 - 12) \\
$m_\Delta$ & 0.595\err{15}{16} (12 -- 16) & 0.593\err{32}{16}
(11 -- 16) &
0.567\err{12}{17} (11 - 16) & 0.559\err{9}{10} (7 - 12) \\
\hline
\end{tabular}
\caption{The estimates of baryon and meson masses using LL, SL,LS
and SS propagators with a Jacobi smearing algorithm comprising 50
iterations at $\kappa_{S} = 0.250$.  Also shown in parentheses is the
time range used in each fit.}
\label{hadron_masses}
\end{center}
\end{table}

As expected, the mass estimates obtained using the SL propagators are
subject to substantially larger errors than those obtained using the
LS propagators, even though the fitting range used is the same.  The
improvement in the determination of the baryon masses through the use
of smeared sources and sinks is substantial.  It is possible that the
study of a larger sample of configurations would lead to a still more
marked improvement through the ability to perform a fit to the full
covariance matrix for the SS correlators over a larger fitting range
than that quoted in Table~\ref{hadron_masses}.

A recent analysis~\cite{daniel92} of the hadron spectrum obtained
using both wall sources and Wuppertal sources suggests that there is a
systematic difference in the baryon masses in the two cases, and that
this difference is particularly noticeable in the determination of the
$\Delta$ mass.  To determine whether such discrepancies exist in our
data, Table~\ref{tab:delta_mass_differences} shows bootstrap analyses
of the $\Delta$ mass differences for the various combinations of
sources and sinks, using the time ranges of Table~\ref{hadron_masses}.
Any discrepancies are at most a $2 \sigma$ effect, and furthermore
appear to depend only on the interpolating field used at the source,
and not on that used at the sink.  Since the expectation values of the
LS and SL correlators should be identical in the limit of an infinite
number of configurations, we attribute the discrepancies to limited
statistics.

\begin{table}
\begin{center}
\begin{tabular}{|c|cccc|}
\hline
& $m_{LL}$ & $m_{SL}$ & $m_{LS}$ & $m_{SS}$ \\
\hline
$m_{LL}$ & \zero{l} & 0.002\err{11}{25} & 0.028\err{15}{17} &
0.036\err{16}{18}\\
$m_{SL}$ & & \zero{l} & 0.026\err{31}{20} & 0.035\err{30}{20} \\
$m_{LS}$ & & & \zero{l} & 0.008\err{19}{18} \\
$m_{SS}$ & & & & \zero{l|} \\
\hline
\end{tabular}

\caption{The entries represent the $\Delta$ mass differences
obtained by subtracting the source-sink combination on the top from
that on the left.}
\label{tab:delta_mass_differences}
\end{center}
\end{table}

\section{Pseudoscalar Decay Constant}
The hope of an improvement in the determination of hadronic
matrix elements provides strong motivation for the use of
smeared operators.  In this section, we take the pseudoscalar decay constant,
$f_\pseudo$, to typify a matrix element that can be computed from
two-point functions.

For a pseudoscalar at rest, the decay constant is related to the matrix
element of the (local) axial vector current by,
\begin{equation}
		\langle 0 | A^4_L(0) | \pseudo \rangle = f_\pseudo m_\pseudo,
\label{defn_fpi}
\end{equation}
where $A_L^\mu$ is the axial operator $\bar{q}\gamma_5\gamma_\mu q$.  For the
following
discussion, we introduce the amplitudes $Z_{L(S)}$ defined by
\begin{equation}
	\langle 0 | P_{L(S)}(0) | \pseudo \rangle = Z_{L(S)} m_\pseudo
\label{defn_z}
\end{equation}
where $P$ is the pseudoscalar operator $\bar{q} \gamma_5 q$, and
the $L$ and $S$ subscripts denote quantities constructed using local and
smeared operators respectively.

We find that the cleanest determination of $f_\pseudo$ using local
sources and sinks~\cite{hadrons_paper} is obtained by fitting to the
ratio
\begin{equation}
\frac{\displaystyle \sum_{\vec{x}} \langle A^4_L(\vec{x},t) P^{\dagger}_L(0)
\rangle}
{\displaystyle \sum_{\vec{x}} \langle P_L(\vec{x},t) P^{\dagger}_L(0)
\rangle} \sim \frac{f_\pseudo}{Z_L} \tanh{m_\pseudo ( T/2 - t)}
\end{equation}
where $T$ is the temporal extent of the lattice, and $Z_L$ and $m_\pseudo$
are determined from the fit to the pseudoscalar correlator.  Using
the fitting range of Table~\ref{hadron_masses}~(12~-~16), we find
\begin{equation}
f_\pseudo = 0.066\mbox{\err{1}{3}},
\end{equation}
where the result is expressed in lattice units.
The determination through the correlator $\langle P_L(\vec{x},t) A^{4
\dagger}_L(0) \rangle$ has larger statistical errors.

{}From propagators computed with both smeared and local sources and
sinks, we can extract $f_\pseudo$ from the ratio
\begin{equation}
\frac{\displaystyle \sum_{\vec{x}} \langle A^4_L(\vec{x},t) P^{\dagger}_S(0)
\rangle}
{\displaystyle \sum_{\vec{x}} \langle P_S(\vec{x},t) P^{\dagger}_S(0)
\rangle} \sim \frac{f_\pseudo}{Z_S} \tanh{m_\pseudo ( T/2 - t)}
\label{f_pi_smeared}
\end{equation}
where $Z_S$ and $m_\pseudo$ are determined from the fit to the appropriate SS
correlator. Using the time range 11 to 16 for the fit to
Equation~(\ref{f_pi_smeared}), and the range 9 to 12 for the fit to
the SS correlator, we obtain
\begin{equation}
f_\pseudo = 0.065\mbox{\err{3}{4}}.
\end{equation}
The determinations of $f_\pseudo$ using the two methods are
consistent.  That we see no improvement using the smeared propagators
may arise from the inability to exploit the full time range in the
correlated fits due to the limited statistical sample.  However, we
note that the statistical error on $f_\pseudo$ using the local
propagators is already small, and that we may be close to the minimum
achievable statistical error.  The improvement in the determination of
the baryon masses through the use of smeared correlators encourages us
to believe that there may be a corresponding reduction in the
uncertainty on baryon matrix elements.

\section{Matrix Correlators}

By studying the matrix of correlators formed from the interpolating
fields constructed using smeared and local sources and sinks, we can
attempt to extract masses for the first excited state.  We begin this
section with a discussion of the case of a general $2 \times 2$ matrix
of correlators.

\subsection{Method}

Consider the two operators, ${\cal O}_a$ and ${\cal O}_b$, having the
same quantum numbers, and a basis of states,
\begin{equation}
	| h_n (\vec{p}) \rangle, \, n = 0, 1, 2, \dots.
\end{equation}
We define the $2 \times 2$ matrix $C(t)$ of timesliced correlators by
\[
	C(t) = \left( \begin{array}{cc}
		c_{aa}(t) & \omega c_{ab}(t) \\
		\omega c_{ba}(t) & \omega^2 c_{bb}(t)
		\end{array} \right)
\]
with
\begin{equation}
	c_{ij}(t) = \sum_{\vec{x}} \langle 0 |
		{\cal O}_i(\vec{x},t) {\cal O}_j^{\dagger}(\vec{0},0)
		| 0 \rangle e^{-i \vec{p}.\vec{x}},
		\label{C_definition}
\end{equation}
where $\omega$ is the arbitrary relative normalisation of the
operators ${\cal O}_b$ and ${\cal O}_a$.
Inserting a complete set of states in Equation~(\ref{C_definition}), we
obtain
\begin{equation}
	C(t) =
		\sum_{n \geq 0} e^{- E_n(\vec{p}) t}
	\left( \begin{array}{cc}
		| a_n |^{2} & a^{\ast}_n b_n \omega \\
		b^{\ast}_n a_n \omega & | b_n |^2 \omega^2
		\end{array}
	\right) \label{SET_OF_STATES}
\end{equation}
where
\begin{equation}
	a(b)_n = \frac{1}{\sqrt{2 E_n(\vec{p})}}
		\langle 0 | {\cal O}_{a(b)} | h_n(\vec{p})\rangle.
\end{equation}
We will now discuss two methods for studying the matrix
$C(t)$~\cite{luscher}.

\begin{enumerate}
\item{\it Diagonalisation of Transfer Matrix}

Consider the eigenvalue equation
\begin{equation}
C(t) u = \lambda(t,t_0) C(t_0) u,
\label{eq:luscher}
\end{equation}
for fixed $t_0 < t$.  If the system comprises only two independent states,
then the eigenvalues, $\lambda_{+}(t,t_0)$ and $\lambda_{-}(t,t_0)$,
of Equation~(\ref{eq:luscher}) are
\begin{eqnarray}
\lambda_{+}(t,t_0) & = & e^{- (t - t_0) E_0} \nonumber \\
\lambda_{-}(t,t_0) & = & e^{- (t - t_0) E_1}.
\label{eq:luscher_eigens}
\end{eqnarray}
The two states are separated exactly, and the coefficients in
Equation~(\ref{eq:luscher_eigens}) grow exponentially with $t_0$.  In
general, where there are more than two states, we expect only two
states to be dominant for sufficiently large $t$; the coefficients of
the contributions of the higher states are suppressed.  Hence ideally
we wish to study $\lambda(t,t_0)$ for $t_0$ as large as possible.
However, the increase in the noise in the data far from the source
generally requires that we choose $t_0$ close to the origin.

We consider a fixed basis of operators by introducing the matrix of
eigenvectors $M(t)$ which diagonalises $C(t_0)^{-1} C(t)$ at $t =
t_0+1$.  In a regime in which effectively we have only two states, the
diagonal elements of the matrix
\begin{equation}
M(t_0+1)^{-1}C(t_0)^{-1} C(t) M(t_0+1)
\end{equation}
are independent of $t_0$, and
equivalent to the eigenvalues of $C(t_0)^{-1} C(t)$.

\item{\it Diagonalisation of $C(t)$}

We can compute the eigenvalues, $\chi_{+}(t,\omega)$ and
$\chi_{-}(t,\omega)$, of $C(t)$ directly, and at large times, $t$,
obtain
\begin{eqnarray}
\chi_{+}(t,\omega) & = & f_{+}(\omega, a_i, b_i) e^{-E_0(\vec{p})t}
 \{ 1 + g_{+}(\omega, a_i, b_i) O(e^{- \Delta E_1(\vec{p}) t}) \}
	\nonumber \\
\chi_{-}(t,\omega) & = & f_{-}(\omega, a_i, b_i)
	e^{- E_1(\vec{p}) t}
\{ 1 + g_{-}(\omega, a_i, b_i) O(e^{ -\Delta \tilde{E}(\vec{p}) t})
\},
\label{eq:sara_eigens}
\end{eqnarray}
where
\begin{eqnarray}
\Delta E_n(\vec{p}) & = & E_n(\vec{p}) - E_0(\vec{p}) \nonumber \\
\Delta \tilde{E}(\vec{p}) & = & \mbox{min}[E_1(\vec{p}) - E_0(\vec{p}),
	E_2(\vec{p}) - E_1(\vec{p})].
\end{eqnarray}

As in method~1, $\chi_{+}(t,\omega)$ and $\chi_{-}(t,\omega)$ are
dominated by the ground state and first excited state respectively.
However, even when there are only two states, corrections to the
leading behaviour of the eigenvalues remain.  The coefficients of
these corrections depend on the arbitrary parameter $\omega$ and,
since we diagonalise $C(t)$ at each timeslice, these coefficients are
not positive definite.  We will exploit the dependence on $\omega$ to
seek cancellations between the contributions from higher states, and
hence extend the plateau region in the effective mass closer to the
source.

Such an approach has clear dangers.  In particular, the effective
masses can approach their asymptotic values either from above or from
below as we vary $\omega$.  However, we observe that for sufficiently
large times the excited state contributions to
Equation~(\ref{eq:sara_eigens}) will be negligible, and thus the
effective masses derived from $\chi_{+}(t,\omega)$ and
$\chi_{-}(t,\omega)$ must be insensitive to $\omega$.

\end{enumerate}

\subsection{Results}

Here we consider ${\cal O}_a$ and ${\cal O}_b$ to be the local
(${\cal O}_L$) and smeared (${\cal O}_S$) interpolating fields
respectively, with $C(t)$ computed at zero spatial momentum.  In
Figure~\ref{lambda_minus_plots}, we show the effective masses of
the first excited states derived from the eigenvalues
$\lambda_{-}(t,t_0)$ for the pseudoscalar, vector, nucleon and
$\Delta$ channels, with $t_0 = 1$.  We find increasing $t_0$ has
no effect on the effective masses except to increase the
statistical error.  Also shown are the effective masses derived
from the smaller diagonal elements of $M(t_0+1)^{-1} C(t_0)^{-1}
C(t) M(t_0+1)$, for $t_0 = 4$; the effective masses for $t_0 = 5$
and $t_0 = 6$ are consistent.  We see that the effective masses
lie on those derived from $\lambda_{-}(t, t_0 = 1)$, and conclude
that two states are effectively dominant for $t_0 \ge 4$.

%
%
\begin{figure}[htbp]
\begin{center}
\leavevmode
\epsfxsize=250pt
  \epsfbox[20 30 620 600]{luscher_pion.ps}
\epsfxsize=250pt
  \epsfbox[20 30 620 600]{luscher_rho.ps}
\end{center}
\begin{center}
\leavevmode
\epsfxsize=250pt
  \epsfbox[20 30 620 600]{luscher_nuc.ps}
\epsfxsize=250pt
  \epsfbox[20 30 620 600]{luscher_delta.ps}
\end{center}
\caption{The effective masses of the first excited states in the pseudoscalar,
vector, nucleon and $\Delta$ channels, derived from
$\lambda_{-}(t,t_0)$, for $t_0 = 1$ (bursts).  Also shown
are the effective masses derived from the smaller diagonal elements of
$M(t_0+1)^{-1} C(t_0)^{-1} C(t) M(t_0+1)$, for $t_0 = 4$ (squares).}
\label{lambda_minus_plots}
\end{figure}

In Figure~\ref{chi_minus_plots}, we show the effective masses of the
first excited states derived from $\chi_{-}(t,\omega)$ for the
pseudoscalar, vector, nucleon and $\Delta$ channels, at $\omega =
10\, \omega_0,\, \omega_0, \mbox{ and } 0.1 \omega_0$, where
\begin{equation}
\omega_0 =
\sqrt{\frac{C_{LL}(t=12)}{C_{SS}(t=12)}}.
\end{equation}
For each channel there is a region in $t$ for which the
effective masses coincide for all $\omega$.  As noted above, this can
be taken as a signal for the first excited state.  The plateau region
in the effective masses appears to extend closer to the source as
$\omega$ decreases.  We attribute this to cancellations between
contributions of the higher excited states to $\chi_{-}(t,\omega)$.
%
%
%
\begin{figure}[htbp]
\begin{center}
\leavevmode
\epsfxsize=250pt
  \epsfbox[20 30 620 600]{omega_pion.ps}
\epsfxsize=250pt
  \epsfbox[20 30 620 600]{omega_rho.ps}
\end{center}
\begin{center}
\leavevmode
\epsfxsize=250pt
  \epsfbox[20 30 620 600]{omega_nuc.ps}
  \epsfxsize=250pt
  \epsfbox[20 30 620 600]{omega_delta.ps}
\end{center}
\caption{
The effective masses of the first excited states in the pseudoscalar,
vector, nucleon and $\Delta$ channels, derived from
$\chi_{-}(t,\omega)$, for $\omega = \omega_0$ (circles), $\omega = 10
\omega_0$ (crosses), and $\omega = 0.1 \omega_0$ (bursts).}
\label{chi_minus_plots}
\end{figure}

Figure~\ref{plot_comparison} shows the effective masses of the first
excited states using method 1, and using method 2 at an approximately
optimal value of $\omega$.  The two methods yield consistent plateaus
in the pseudoscalar, vector and nucleon channels, and using method~2
we obtain
\begin{eqnarray}
\tilde{m}_\pseudo (5 - 8) & = &
0.93\mbox{\err{6}{13}} \nonumber \\
\tilde{m}_V (5 - 8) & = & 0.90\mbox{\err{3}{4}}
\nonumber \\
\tilde{m}_{N} (6 - 9) & = & 1.02\mbox{\err{6}{6}}
\label{eq:excited_mass_fits}
\end{eqnarray}
where the fitting range is shown in brackets.  Our fitting procedure
is as follows.  We pick a value of $\omega$ which provides the largest
plateau region in the effective mass of the first excited state.  We
then perform a correlated fit to the eigenvalue $\chi_{-}(t, \omega)$,
using a fitting range such that the fit is insensitive, within
statistical errors, to the addition of the timeslice closer to the
source.  We estimate the systematic errors by looking at the variation
in the central value of the mass as we vary $\omega$ about the optimal
value, keeping the time range fixed; we find the systematic errors to
be of order 5\%.

There is no clear plateau in the effective mass in the $\Delta$
channel using method~1. However, using method~2 we obtain
\begin{equation}
\tilde{m}_\Delta (6 - 9) =
 1.02\mbox{\err{5}{3}}.
\end{equation}
%

%
%
\begin{figure}[htbp]
\begin{center}
\leavevmode
\epsfxsize=250pt
  \epsfbox[20 30 620 600]{comp_pion.ps}
  \epsfxsize=250pt
\epsfbox[20 30 620 600]{comp_rho.ps}
\end{center}
\begin{center}
\leavevmode
\epsfxsize=250pt
  \epsfbox[20 30 620 600]{comp_nuc.ps}
\epsfxsize=250pt
  \epsfbox[20 30 620 600]{comp_delta.ps}
\end{center}
\caption{The effective masses of the first excited states in the
pseudoscalar, vector, nucleon and $\Delta$ channels derived from
$\lambda_{-}(t,t_0)$ at $t_0 = 1$ (circles), and from
$\chi_{-}(t,\omega)$ at $\omega = 0.1\omega_0$ (diamonds).}
\label{plot_comparison}
\end{figure}

We find that the masses of the first excited states are more than
twice those of the corresponding ground states.  The quark mass
employed in our investigation is somewhat less than that of the
strange quark.  The measured ratios are therefore considerably larger
than the corresponding experimental ratios for hadrons consisting
entirely of strange quarks, e.g.  $m_{\Omega(2250)^{-}} /
m_{\Omega^{-}}
\simeq 1.35$, $m_{\phi(1680)} / m_{\phi(1020)} \simeq 1.65$.  If this
is disappointing, it should be noted that all our computed values of
the first excited state masses are $O(1)$.  Since the mass differences
between these states and the corresponding ground states arise from
radial excitations, the discretisation errors may be substantial.
Furthermore, unexpectedly large values for the masses of the excited
states have been observed in other simulations~\cite{marinari92}.  We
conclude this section by remarking that an investigation of the larger
eigenvalue, $\lambda_{+}(t, t_0)$, of method 1 yields similar results
for the ground state masses to those obtained from the analysis of the
SS correlators.

\section{Conclusions}

We have demonstrated a gauge-invariant smearing algorithm that
provides a considerable improvement in the determination of baryon
masses. Though we see some discrepancy between the baryon masses
obtained using local sources and those obtained using smeared sources,
it is very small, and we attribute it to the limited statistics.

An investigation of the $2 \times 2$ matrices of correlators formed
from the LL, SL, LS and SS propagators allows us to extract the masses
of the first excited states.  We have explored two methods that yield
consistent results.  It would be interesting to extend this analysis
to a larger basis of operators; such an investigation should lead to
appreciably better discrimination of the lowest two states.

\section{Acknowledgements}

This research is supported by the UK Science and Engineering Research
Council under grants GR/G~32779, GR/H~01069 and GR/H~49191, by the
University of Edinburgh and by Meiko Limited.	We are grateful to
Mike Brown of Edinburgh University Computing Service, and to Arthur
Trew and Paul Adams of EPCC, for provision and maintenance of service
on the Meiko i860 Computing Surface and the Thinking Machines CM-200.
CTS and ADS acknowledge the support of the UK Science and Engineering
Research Council through the award of Senior and Personal Research
Fellowships respectively.

\end{document}